\documentstyle[aps,prb,epsfig,multicol,color]{revtex}

\begin{document}
\draft

\title{ Anomalous specific heat jump in the heavy fermion superconductor CeCoIn$_5$}

\author{Yunkyu Bang$^{*}$, A.V. Balatsky}

\address {Theoretical Division Los Alamos National Laboratory, Los Alamos, New Mexico \
87545 }

\date{\today}
\maketitle

\begin{abstract}
We study the anomalously large specific heat jump and its systematic change
with pressure in CeCoIn$_5$ superconductor. Starting with the general free
energy functional of the superconductor for a coupled electron boson system, we
derived the analytic result of the specific heat jump of the strong coupling
superconductivity occurring in the coupled electron boson system. Then using
the two component spin-fermion model we calculate the specific heat coefficient
$C(T)/T$ both for the normal and superconducting states and show a good
agreement with the experiment of CeCoIn$_5$.
Our result also clearly demonstrated that the specific heat coefficient
$C(T)/T$ of a coupled electron boson system can be freely interpreted as a
renormalization  either of the electronic or of the bosonic degrees of freedom.
\end{abstract}



\begin{multicols}{2}

The superconductivity (SC) in CeMIn$_5$ (M=Co, Rh, Ir) heavy fermion (HF)
series compounds \cite{Hegger,Petrovic} provides valuable information about the
superconductivity nearby a possible quantum critical point (QCP) due to the
excellent tunability with pressure and magnetic fields\cite{Co QCP}. Besides a
complicated phase diagram due to a competetion/interplay between the magnetism
and superconductivity, the superconducting properties as well as the normal
state ones around the QCP exhibit various anomalous behaviors such as non-Fermi
liquid (NFL) power law of temperature in the resistivity, pseudogap behavior
above the superconducting critical temperature, the continuous evolving from
the second order to the first order phase transition of the superconductivity
with magnetic fields, etc\cite{Co QCP}.

In a recent paper\cite{Bang}
 we proposed the two-component spin-fermion model as a generic pairing
mechanism of this series of compounds CeMIn$_5$. In this work, it is shown that
some of the standard features of the d-wave superconductivity mediated by
magnetic fluctuations becomes strongly modified in the vicinity of the magnetic
critical point. Indeed it is very natural to expect that such a strongly
coupled fermion-boson superconductivity shows various anomalies deviating from
the standard weak or strong coupling theory of superconductivity\cite{Abanov}.

In this paper we study  anomalous property of the specific heat coefficient
$C(T)/T$ in CeCoIn$_5$. Sparn et al.\cite{Sparn} measured $C(T)/T$ of
CeCoIn$_5$ with varying pressure and also with applying magnetic fields. The
key findings are that (1) suppressing the superconductivity with 8 Tesla fields
$C(T)/T$ show $\ln T$ divergence down to 0.4 K; (2) without the magnetic fields
the same sample becomes superconducting at 2.3 K with the specific heat jump
ratio $\Delta C(T)/C(T) \sim 5$, which is a huge jump considering $C(T_c)/T_c
\sim 400mJ/K^2 mol$; (3) with applying pressure $T_c$ gradually increases to
2.8 K at the pressure of 15 kbar but the jump ratio $\Delta C(T)/C(T)$  drops
drastically. We think  {\it this peculiar behavior is another manifestation of
the critical magnetic fluctuations mediated superconductivity.} The huge
specific heat jump, indicating the steep drop of the entropy below $T_c$, is
actually consistent with the spin-fermion model if we consider that the
critical magnetic fluctuations in this model is generated by the fermionic
particle-hole excitations, which should be gapped below $T_c$. This means that
below $T_c$ not only the fermions suffer the entropy drop by developing pairing
correlation but also the spin fluctuations should suffer the entropy drop,
which add up to make a larger specific heat jump. As will be shown below,
however, this phenomena is actually very general so that it should occur even
in the electron-phonon mediated superconductors. The reason why it is
phenomenal in CeCoIn$_5$ is simply because the coupling of fermion-boson is
strong and in particular the resulting bosonic mode is near critical. There is
a recent work \cite{Kos} addressing the same problem as in this paper, but
proposing a different mechanism for the enhanced $\Delta C(T)/C(T)$.

We start with the general free energy functional of the two component
spin-fermion model, which is nothing but the same form as the one for
electron-phonon model\cite{Luttinger-Ward,Eliashberg,Bardeen-Stephen}.

\begin{eqnarray}
 \Omega_{total} &=& \Omega_{fer}+ \Omega_{spin} + \Omega_{int} \\
&=& -\frac{T}{V} \sum_P [\ln (-\phi(P)) + 2 \Sigma_{1}(P)G(P) - 2
\Sigma_{2}(P)F(P)] \nonumber \\
 && +(\frac{3}{2} \frac{T}{V}) \sum_q [\ln(-D^{-1}(q)+\pi(q) D(q)]  \nonumber \\
&& + (T/V)^2 \sum_{P,P^{'}} \alpha_{P-P^{'}} ^2 [G(P)D(P-P^{'})G(P^{'})
\nonumber \\ && -F(P)D(P-P^{'})F(-P^{'}) ]
 \end{eqnarray}

\noindent
 where
$P=(\vec{p},i \omega_{n}), q=(\vec{q},i \Omega_{n});  \omega_{n}=\pi T (2n+1),
\Omega_{n}=\pi T (2n)$, and
%
$G(P)=(i \omega_{n}+\epsilon_{p}+\Sigma_{1}(-P))/\phi(P)$,
%
$F(P)=- \Sigma_{2}(-P)/\phi(P)$,
%
$\phi(P)= (i \omega_{n}-\epsilon_{p}-\Sigma_{1}(P))(i
\omega_{n}+\epsilon_{p}+\Sigma_{1}(-P)) - |\Sigma_{2}(P)|^{2}$, respectively.
%
Finally, the dressed spin propagator is written as $D^{-1}(q)=D_{0}
^{-1}(q)-\pi(q)$.
%
$\Omega_{fer}, \Omega_{spin}$, and  $\Omega_{int}$ are just apparent separation
of the total free energy but the literal meaning of them is not necessarily
correct as will be clear later.  The notations are standard
\cite{Bardeen-Stephen}, and the one important difference is that $D_{0} (q)$ is
not explicitly defined in the spin-fermion model in contrast to the
electron-phonon case, but only  the total spin propagator is defined
phenomenologically as follows.

\begin{equation}
D(q)=\frac{D_{0}}{[1-D_{0}\pi(q)]}=\frac{D_{0}}{I(T)+A(\vec{q}-\vec{Q})^2+
|\Omega_n|/\Gamma}.
\end{equation}

\noindent where $I(T)$ defines the distance from the magnetic quantum critical
point (QCP) and $A$ and $\Gamma$ are the coefficients determined by the Fermi
liquid parameters of the fermion sector \cite{Hasegawa}. The effective spin
correlation length and the relaxational energy scale are $\xi^2=A/I(T)$ and
$\omega_{sf}=\Gamma \cdot I(T)$, respectively, and the momenta $\vec{q}$ and $
\vec{Q}$ are assumed to be in the two dimensions\cite{Rosch}. We take $\Gamma$
as the unit energy scale in our numerical calculations. The bare spin
propagator $D_{0}$ is assumed to be trivial regarding the critical low energy
behavior and taken to be a constant. As usual, the above free energy functional
is constructed in such way that it is stationary with respect to the variations
of $\Sigma_{1,2}$ and $\pi$ a la Luttinger-Ward\cite{Luttinger-Ward} with the
following definitions. This property of the functional is very important and
utilized critically in our paper.

\begin{eqnarray}
\Sigma_{1}(P) &=&\frac{T}{V} \sum_{P^{'}} \alpha_{P-P^{'}} ^{2} G(P^{'})
D(P-P^{'}),
\\ \Sigma_{2}(P)&=&\frac{T}{V} \sum_{P^{'}} \alpha_{P-P^{'}} ^{2} F(P^{'}) D(P-P^{'}),
\\ \pi(P)& = & - \frac{2}{3}  \frac{T}{V} \cdot \nonumber \\
& & \sum_{P} \alpha_{q} ^2 [G(P+q)G(P)-F(P+q)F(-P)].
\end{eqnarray}

\noindent Now following Ref[\cite{Bardeen-Stephen}], neglecting the momentum
dependence of $\Sigma_{1,2}$ and using the definitions, $
%
\Sigma_{1}(P)= i \omega_{n} (1- Z(\omega_{n})),$ and $
\Delta(\omega_{n})=\Sigma_{2}(P)/Z(\omega_{n}),$
%
we obtain after the momentum integration  the following.

\begin{eqnarray}
\Omega_{fer}+ \Omega_{int}  = -T \cdot N(0) \cdot \pi  \nonumber \\
\sum_{\omega_{n}} [ (Z(\omega_{n})+1) \sqrt{\Delta^2-(i \omega_{n})^2} -
\frac{\Delta^2}{\sqrt{\Delta^2-(i \omega_{n})^2}}]
\end{eqnarray}

\noindent Note that the above equation didn't include the $\Omega_{spin}$ term
yet \cite{S_boson}. But due to the stationary property of the total free energy
functional as mentioned above, the specific heat jump across the
superconducting transition with the changes of  $\Sigma_{1} ^{(n)} \rightarrow
\Sigma_{1} ^{(s)}$ and $\pi^{(n)} \rightarrow \pi^{(s)}$  is solely determined
by Eq.(7) with $Z(\omega_{n})= Z^{(n)}(\omega_{n})$ at $T=T_{c}$.

The Eq(7) can be written after the Matsubara frequency summation using the
contour integration as

\begin{eqnarray}
\Omega_{fer}+ \Omega_{int}  =  N(0) \int_{0} ^{\infty} d \epsilon [(Z+1) \omega
- Z \frac{\Delta^2}{\omega}] \cdot \tanh(\frac{\omega}{2 T}),
\end{eqnarray}

\noindent where $\omega^2=\epsilon^2+\Delta^2$ and the specific heat is
calculated at $T=T_c$ as

\begin{equation}
C(T)/T=(Z+1) N(0) \frac{\pi^2}{3}+\frac{Z N(0)}{T_c}(-\frac{\partial \Delta^2
}{\partial T})
\end{equation}

 This is our key result in this paper and we can read the specific heat jump
ratio as $\frac{\Delta C(T)}{C(T)}=\frac{Z}{Z+1} \cdot \frac{-\frac{\partial
\Delta^2 }{\partial T}/ T_c}{\pi^2/3}$ \cite{S_boson}. We see from this formula that the jump
ratio is modified with  the normal self-energy renormalization factor
$\frac{Z}{Z+1}$, which is $\frac{1}{2}$ in the BCS limit with $Z=1$ and can
have a maximum value $1$ in the limit of $Z(T) \rightarrow \infty$. Apparently
$\frac{Z}{Z+1} \rightarrow 1$  is the proper limit for CeCoIn$_5$ and some
other heavy fermion compounds exhibiting a large $C(T)/T$ at low temperature.
We summarize the consequences of this formula: (1) The jump ratio at most can
be doubled to the BCS ratio assuming $\frac{\partial \Delta^2 }{\partial T}$ is
the same as the BCS limit, i.e. $9.42 T_c$; (2) Apparently the steeper slope of
$\Delta(T)$ can contribute an extra enhancement to the jump ratio. And since
our formula is an exact result, this formula can be utilized to estimate the
temperature slope of $\Delta(T)$ at $T_c$ from the experimental $\frac{\Delta
C(T)}{C(T)}$ in the limit of  $\frac{Z}{Z+1} \rightarrow 1$;
 (3) Finally, the interpretation of
the large $C(T)/T$ and its enhanced jump ratio from the heavily renormalized
fermionic quasiparticle due to the scattering from the magnetic fluctuations
seems to be legitimate, if we can neglect the contribution from
$\Omega_{boson}$  \cite{S_boson}.


Now in order to calculate $C_{total}(T)$ above and below $T_c$, we need to
calculate $\Sigma^{(n),(s)} _{1,2}$ with a given $D(q)^{(n),(s)}$.  But we
found a more convenient way to proceed as follows.
 Adding $\Omega_{int}$ term to  $\Omega_{spin}$  exactly cancels
the cumbersome term $\pi(q) D(q)$ as

\begin{equation}
\Omega_{spin} + \Omega_{int} = (\frac{3}{2} T/V) \sum_q [\ln(-D^{-1}(q)].
\end{equation}

 Furthermore,  $\Omega_{fer} $ term alone,
after the same manipulations used for  Eq.(7) and Eq.(8), becomes

\begin{eqnarray}
\Omega_{fer}& = & -T \cdot N(0) \cdot \pi   \sum_{\omega_{n}} [ 2
\sqrt{\Delta^2-(i \omega_{n})^2} - 2  \frac{\Delta^2}{\sqrt{\Delta^2-(i
\omega_{n})^2}}] \nonumber \\ & = & N(0) \int_{0} ^{\infty} d \epsilon ~ 2
\omega
 \cdot \tanh(\frac{\omega}{2 T}).
\end{eqnarray}

A different organization greatly simplifies  $\Omega_{fer}$ so that all the
renormalization effect due to the interaction miraculously drops  and
furthermore the explicit $\Delta(T)$ dependent term all cancels out in the
final result of Eq.(11). This amazing result is actually the consequence of the
Eliashberg free energy functional built with the stationary property with
respect to the variations of the self-energies. Considering that  the
Eliashberg free energy functional is not the exact free energy functional but a
consistent approximate functional equivalent to the self-consistent Born
approximation for the fermionic part and the random phase approximation (RPA)
for the bosonic part, the exactness of Eq.(9) and Eq.(11) has a limited
meaning. Nevertheless, besides the accuracy of this functional the insightful
reason for this result is the equivalent relation of the following.

\begin{equation}
\frac{T}{V} \sum_P [ \Sigma_{1}(P)G(P) -  \Sigma_{2}(P)F(P)]
 = - (\frac{3}{2} \frac{T}{V}) \sum_q [\pi(q) D(q)].
\end{equation}

From the technical viewpoint the above relation is trivial if we consider any
linked-cluster diagram obtained from the free energy expansion with $H_{int}$
\cite{Luttinger-Ward}. However the physical interpretation is rather revealing.
Namely, the interacting free energy $\Delta
\Omega=\Omega_{total}-\Omega_{bare}$ and its derived entropy $\Delta S$ all
comes from the expansion of $H_{int}$. Then depending on how to view or
organize the same diagrams, we can view all interaction effect as the
renormalization of either the fermion sector or the bosonic sector
\cite{Kotliar}.
While theoretically we have this freedom, in experiments   the experimental
probe itself determines how to measure the contribution of $H_{int}$ through
either fermionic or bosonic degree of freedom. However the thermodynamic
measurement like the specific heat measures simultaneously the both degrees of
freedom  and it is hard to know how much contribution comes from which degrees
of freedom. Our analytic result  clearly demonstrated that the specific heat
coefficient $C(T)/T$ of a coupled fermion-boson system can be freely
interpreted as a renormalization either of the electronic or of the bosonic
degree of freedom, but not for both.
This point should have a far-reaching consequence to the heavy fermion
experiments and its related interpretations. This important aspect will be
dealt in the separate publication.

Now let us return to the calculation for the total $C(T)_{tot}$. From Eq.(10)
we can calculate the entropy $S(T)_{spin+int}$ with the phenomenological $D(q)$
given in Eq.(3) as follows.

\begin{eqnarray}
S_{spin}+S_{int}= \sum_{q} \int_{0} ^{\Lambda_{\omega}} \frac{3 d \omega}{4\pi
} [\frac{1}{\sinh (\frac{\omega}{2 T})} ]^2   \nonumber \\ \cdot
\frac{\omega}{T^2} \cdot \arctan \frac{\omega/\Gamma}{(I_{0}+a T)+ A
(\vec{q}-\vec{Q})^2}.
\end{eqnarray}

In real calculation, we need to fix  parameters. Our unit energy scale is
$\Gamma=1$, and we take $\Lambda_{\omega}=3$, $A
(\vec{q}^{max}-\vec{Q})^2=\frac{\omega_{q} ^{max}}{\Gamma}=1.0$, and $a=1.0$.
%
For the fermionic part  $\Omega_{fer}$, it is just the same form as the
non-interacting fermion free energy except $\omega =\sqrt{\epsilon_{k}
^2+\Delta^2 (T)}$, and therefore it develops some structure below $T_c$ but no
discontinuity at $T_c$. Finally to fix the relative magnitude of
$\Omega_{spin+int}$  to  $\Omega_{fer}$  we need to determine the spin density
of states $N_{s}$ defined as $N_s \int_{0} ^{\Lambda_{\omega}} =\frac{1}{V}
\sum_{q} ^{q_{max}}$. This can be determined using Eq.(12) at normal state,
which gives the following relation

\begin{eqnarray}
\frac{3}{2} T N_{s} \sum_{\Omega_{n}} |\Omega_{n}| \ln [ \frac{ I(T)+\omega_{q}
^{max}/\Gamma+|\Omega_{n}|/\Gamma} { I(T)+|\Omega_{n}|/\Gamma}] = \nonumber
\\ T N(0) \pi \sum_{\omega_{n}} (Z^{(n)}-1) |\omega_{n}|.
\end{eqnarray}

We assume $\pi_{n}=-\frac{|\Omega_{n}|}{D_0 \Gamma}$ in the derivation, but the
above relation is not useful since both side of the  equation diverges unless
we introduce high frequency cut-off for $\Omega_{n}$ and $\omega_{n}$. Since
the above relation should hold for a variation of $\delta \pi_{n}$, we derive
another equivalent relation at $T=0$ as

\begin{eqnarray}
\frac{3}{2 \pi}  N_{s} \int_{0} ^{\Lambda_{\omega}} d \Omega [
\frac{-\Lambda_{\omega}}{[ I(T)+\Lambda_{\omega}+\Omega] [I(T)+\Omega]}]
\Omega^2 \nonumber \\
 = N(0) \int_{0} ^{\infty} d \omega ~~ \omega \cdot \frac{
[ Z^{(n)} (\omega; \pi_{n} +\delta \pi_{n}) - Z^{(n)} (\omega; \pi_{n}) ]}{
\delta \pi_{n}}.
\end{eqnarray}

A recent paper by Ref.[\cite{Haslinger}] also use the similar procedure to fix
the relation between $N(0)$ and $N_{s}$. Our numerical result gives $N_{s}
\cong 1.5 \frac{D_0 \alpha^2}{\pi} N(0)$. Now we are ready to calculate
$C_{tot}(T)$ using Eq.(11) and Eq.(13). For the calculation below $T_c$ we need
to know the temperature dependence of the gap function $\Delta(T)$ and $\pi_{s}
(\Delta(T))$. For the gap function we use the generalized BCS form
$\Delta(T)=\Delta_0 \tanh (a \sqrt{T_c/T -1})$, where $a=1.764$ and $\Delta_0=
1.74 T_c$ for BCS limit, but for strong coupling superconductor these values
can deviate largely from the BCS limit and we take them as parameters. Now much
unclear part is the behavior of $\pi_{s} (T)$ below $T_c$.
Qualitatively it should be cut off for $\Omega \leq 2 \Delta_0$ at $T=0$ and
smoothly recover to the normal state form approaching $T \rightarrow T_c$. The
leading expansion of  $\pi_{s} (T)$ in $\Delta^2 (T)$  gives $\pi_{s}=\pi_{n}
\exp (- \Delta^2(T)/[\omega^2+T^2])$; but for a larger value of $\Delta(T)$ for
$T \rightarrow 0$ this form should not be very correct.

The numerical results are shown in Fig.(1) and Fig.(2). In Fig.(1) the specific
heat coefficient $C(T)/T$ calculated from Eq.(13) are shown in unit of
$\frac{D_0 \alpha^2}{\pi \Gamma} N(0)$. For our parameter choice producing $Z
>> 1$ (For this we need the critical spin fluctuations ($I_0 << 1$) as well as  the
effective coupling $\frac{D_0 \alpha^2}{\pi}$ of  $O(1)$), the contribution of
$C(T)/T$ from $\Omega_{fer}$ Eq.(11) is negligibly small. Therefore we didn't
add it to $C(T)/T$ in Fig(1) but show it separately in Fig.(2) demonstrating
its qualitative features.  In Fig.(1) the solid line is for the normal state
$C(T)/T$ with $I_0=0$ (the spin fluctuations is at QCP), it indeed displays
$\ln T$ divergence with decreasing temperature. With SC transition at T=0.2
(open square symbol) it shows the jump ratio $\Delta C(T)/C(T) \sim 5$
 with the choice of parameters \cite{comment2}. Then increasing $I_0$, the $\ln
T$ divergence is quickly suppressed and at the same time the specific heat jump
ratio drops with increasing $T_c$ ($T_c$ is increased by hand and the
temperature slope of $\Delta(T)$ is reduced accordingly as
$\frac{\Delta^{'}(T_{c1})}{\Delta^{'}(T_{c2})} = \frac{T_{c2}}{T_{c1}}$). With
this rather crude phenomenological  choice of parameters our numerical results
reproduce the experimental features of $C(T)/T$ in CeCoIn$_5$ qualitatively as
well as quantitatively.

In summary, starting with the general free energy functional for the coupled
fermion-boson system, we  derived an analytic formula of $\Delta C(T)/C(T)$ for
the general strong coupling superconductor.
Then we calculate $C(T)/T$ for the spin-fermion model and show that the salient
features of  $C(T)/T$ of CeCoIn$_5$ for both normal and superconducting states
are successfully explained  including its anomalous jump ratio $\Delta
C(T)/C(T)$ and the progressive reduction of it with increasing $T_c$. Also with
different organization of the Eliashberg free energy functional, which lumps
all the interaction effect either into  the bosonic or into the  fermionic
degrees of freedom, we clearly demonstrated that {\it the effect of the
interaction in the total free energy or entropy of the coupled fermion-boson
system can be freely viewed as the renormalization of either the fermion sector
or  the boson sector.} Finally, our results strongly support the idea that the
two dimensional critical magnetic fluctuations plays an essential role in
CeMIn$_5$ HF compounds in producing a large and strongly temperature dependent
$C(T)/T$ and the SC pairing itself.


We are grateful to  J.D. Thompson, J.L. Sarrao, Ar. Abanov, and C. M. Varma for
useful discussions. This work was supported  by US DoE. Y.B. was also partially
supported by the Korean Science and Engineering Foundation (KOSEF) through the
Center for Strongly Correlated Materials Research (CSCMR) (2001) and through
the Grant No. 1999-2-114-005-5.

\begin{figure}
\epsfig{figure=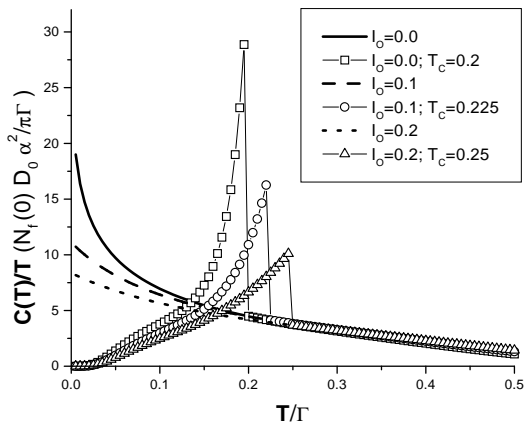,width=1.0\linewidth} \caption{ $(C(T)/T)_{spin+int}$
from $\Omega_{spin} + \Omega_{int}$ contribution in unit of $D_0 \alpha^2/ \pi
\Gamma N_f (0)$ \label{fig1}}
\end{figure}

\begin{figure}
\epsfig{figure=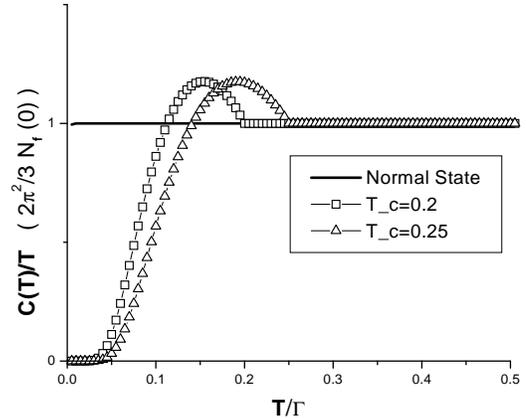,width=1.0\linewidth} \caption{ $(C(T)/T)_{fer}$ from
$\Omega_{fer}$ contribution in unit of $(2 \pi^2/3)  N_f (0)$  \label{fig2}}
\end{figure}

\end{multicols}


\begin{references}
\bibitem{S_boson}
In the electron-phonon case, the term $\Omega_{boson}$ (in this paper
$\Omega_{spin}$)  is neglected with the argument that the change $(\pi_s
-\pi_n) \sim \Delta^2/E_{f}^2$ is small\cite{Bardeen-Stephen}. But this
argument is  wrong because the change of the free energy above and below $T_c$
is always $O(\Delta^2/E_{f}^2)$ and this small change is what makes the
superconducting transition. As shown by Chester\cite{Chester} even before the
BCS theory, the actual change of phonon kinetic energy is just the same order
of magnitude as the electronic part. The correct reason why still the term
$\Omega_{boson}$ can be neglected is that the contribution of the self-energy
correction $\pi_{s,n}$ in $\Omega_{boson}$ exactly cancels between
$\ln(-D^{-1}(q))$ and $\pi(q) D(q)$ terms for the small frequency limit. Then
the actual contribution of $\Omega_{boson}$ is just the one of the
unrenormalized bare boson (or phonon) propagator, so it can be dropped.
\bibitem{comment2}
In order to compare with experiment, we should count the gap anisotropy which
reduces the specific heat jump by the factor $<\Delta^2(k)>^2/
<\Delta^4(k)>$\cite{Mishonov} (for d-wave gap, this value is 0.666.). In
Fig.(1), we just multiplies 0.666 to the results below $T_c$ for a better
comparison with the experiment, but this artificial procedure ruins the entropy
conservation.
\end{references}
\end{document}